\documentclass[aps,groupedaddress,showpacs,floatfix]{revtex4}
\usepackage{amssymb}
\usepackage{amsmath}
\usepackage{graphicx}
\begin{document}

\thispagestyle{empty}

\title{ One dimensional directed polymer "memory model"}

\author{Victor Dotsenko}

\affiliation{LPTMC, Sorbonne Universit\'e, Paris, France}

\date{\today}

\begin{abstract}
In this paper  we propose very simple statistical "memory model" 
of one-dimensional directed polymers which is capable to store and retrieve 
a given random quenched trajectory. The model is defined in terms of the 
elastic string Hamiltonian with the local attractive potential between
the dynamic and the quenched random strings. The average 
overlap between them is calculated  as a function of the 
temperature and the strength of the attractive potential.

\end{abstract}


\maketitle

\section{Introduction}

Usually the statistical properties of one-dimensional directed polymers
are studied in the context of their free energy fluctuations caused by the presence of 
quenched random potential (see e.g. \cite{KPZ,hh_zhang_95,Corwin,Borodin,Rev}).
Here I would like to propose another type of the directed polymer model 
in which quenched disorder is present in a form of a given random trajectory.
Namely, we will consider  the model of directed polymers defined in terms
of an elastic string $\phi(\tau)$ directed along the $\tau$-axes within an interval $[0,t]$ 
in the presence of another {\it quenched random} string $\varphi(\tau)$
with an attractive interaction potential $U\bigl[\phi(\tau)-\varphi(\tau)\bigr]$ between them.
Both strings have the same boundary constraints $\phi(0) = \phi(t) = 0$
and $\varphi(0) = \varphi(t) = 0$. For a given quenched trajectory $\varphi(\tau)$ 
the energy of the string $\phi(\tau)$ is
\begin{equation}
\label{1}
H[\phi(\tau); \varphi(\tau)] = \int_{0}^{t} d\tau
\Bigl\{\frac{1}{2} \bigl[\partial_\tau \phi(\tau)\bigr]^2
+ U\bigl[\phi(\tau)-\varphi(\tau)\bigr]\Bigr\}
\end{equation}
The interaction potential $U(\phi)$ is supposed to be short-ranged:
\begin{equation}
\label{2}
U(\phi) \; = \; -\frac{u}{\epsilon \sqrt{2\pi}} \, \exp\Bigl\{-\frac{1}{2\epsilon^{2}} \phi^{2}\Bigr\}
\end{equation}
where positive parameter $u$ defines the strength of the interaction and $\epsilon$ is the typical 
interaction distance of the quenched and the dynamical trajectories.
The (normalized) probability distribution function ${\cal P}\bigl[\varphi(\tau)\bigr]$  
of the quenched random trajectories $\varphi(\tau)$ (such that $\varphi(0) = \varphi(t) = 0$)
is taken in the Gaussian form
\begin{equation}
\label{3}
{\cal P}\bigl[\varphi(\tau)\bigr] \; = \; \sqrt{\frac{2\pi t}{\gamma}} \,
\exp\biggl\{
-\frac{1}{2} \gamma \int_{0}^{t} d\tau 
\;
\bigl(\partial \varphi(\tau)\bigr)^{2}
\biggr\}
\end{equation}
where $\gamma$ is the elasticity parameter.

For a given quenched  
trajectory $\varphi(\tau)$  the partition function of this system is
\begin{equation}
\label{4}
Z\bigl[\varphi(\tau)\bigr] \; = \; 
\int_{\phi(0)=0}^{\phi(t)=0} {\cal D}\phi(\tau) \,
\exp\biggl\{
- \beta \, \int_{0}^{t} d\tau 
\Bigl[
\frac{1}{2} \bigl(\partial_\tau \phi(\tau)\bigr)^2 \; + \; 
 U\bigl[\phi(\tau)-\varphi(\tau)\bigr]
\Bigr]
\biggr\}
\end{equation}
where $\beta = 1/T$ is the inverse temperature, and the integration goes over all (continuous)
trajectories with the zero boundary conditions. The integration measure is chosen such that
\begin{equation}
\label{5}
\int_{\phi(0)=0}^{\phi(t)=x} {\cal D}\phi(\tau) \,
\exp
\biggl\{
-\frac{1}{2} \beta \, \int_{0}^{t} d\tau \; \bigl(\partial_\tau \phi(\tau)\bigr)^2 
\biggr\}
\; = \; 
\sqrt{\frac{\beta}{2\pi \, t}} \,
\exp\Bigl\{ -\frac{\beta}{2 t} \, x^{2} \Bigr\}
\end{equation}
The physical (selfaveraging) free energy of this system is defined in the usual way
\begin{equation}
\label{6}
F(T, u, m) \; = \; -\frac{1}{\beta} \, \overline{\log\Bigl(Z\bigl[\varphi(\tau)\bigr]\Bigr)}
\end{equation}
where $\overline{ (...)}$ denotes
the averaging over random trajectories $\varphi(\tau)$ with the probability
distribution (\ref{3}).

We will study this system in the thermodynamic limit $t \to \infty$.
The final goal of this study is to calculate
the mean square difference between $\phi(\tau)$ and $\varphi(\tau)$, namely
\begin{equation}
\label{7}
q \; = \; \lim_{t\to\infty} \,\frac{1}{t} \int_{0}^{t} d\tau \, 
\overline{ \big\langle \bigl(\phi(\tau) - \varphi(\tau)\bigr)^{2} \big\rangle }
\end{equation}
where $\langle ... \rangle$ denotes the thermal average over the string trajectories
$\phi(\tau)$ defined by the Hamiltonian (\ref{1}). Then the "order parameter" which 
describe the quality of retrieval  of a given quenched random trajectory  $\varphi(\tau)$
by the dynamic string $\phi(\tau)$ can be defined as
\begin{equation}
\label{8}
m \; = \; \frac{1}{1 \, + \, q}
\end{equation}
In the case of perfect retrieval, when $q \to 0$, we get $m \to 1$, while in the case
of no retrieval, when $q \to \infty$, we get $m \to 0$

In this paper we will consider two limit cases cases. 
First (Section II), supposing that the trajectories on average are close to each other,  
such that  $q \ll \epsilon^{2}$,  
the interaction potential can be approximated as
\begin{equation}
\label{9}
U(\phi-\varphi) \; \simeq \; -\frac{u}{\epsilon \sqrt{2\pi}} \; + \; 
\frac{u}{2 \epsilon^{3} \sqrt{2\pi}} \, \bigl(\phi-\varphi\bigr)^{2}
\end{equation}
Second (Section III), supposing that the trajectories on average are far 
far from each other, such that $q \gg \epsilon^{2}$, the interaction  potential 
can be approximated as
\begin{equation}
\label{10}
U(\phi-\varphi) \; \simeq \; - u \, \delta (\phi-\varphi)
\end{equation}
It will be shown that in both cases the retrieval parameter $m$, eq.(\ref{8}), 
can be computed explicitly exhibiting smooth dependence on $T$ and $u$. 
Finally, Section IV is devoted to the brief discussion of the obtained results.

\section{Gaussian approximation} 

In the Gaussian approximation, eq.(\ref{9}), (neglecting irrelevant constant   
term $-u/\epsilon \sqrt{2\pi}$) the Hamiltonian of the considered system,
eq.(\ref{1}), becomes
\begin{equation}
\label{11}
H[\phi(\tau); \varphi(\tau)] = \int_{0}^{t} d\tau
\Bigl\{\frac{1}{2} \bigl[\partial_\tau \phi(\tau)\bigr]^2
\; + \; \frac{1}{2} \lambda \, \bigl(\phi(\tau)-\varphi(\tau)\bigr)^{2} \Bigr\};
\end{equation}
where
\begin{equation}
\label{12}
\lambda \; = \; \frac{u}{\epsilon^{3} \sqrt{2\pi}}
\end{equation}
In terms of the standard replica approach the average free energy, eq.(\ref{6}), 
can be computed as
\begin{equation}
\label{13}
F \; = \; - \lim_{N\to 0} \, \frac{1}{\beta N} \, \log\bigl(Z_{N}\bigr)
\end{equation}
where
\begin{eqnarray}
\nonumber
Z_{N} &\equiv& \overline{\Bigl(Z[\varphi(\tau)]\Bigr)^{N}} \; = \; 
\\
\nonumber
\\
&=& 
\int_{\varphi(0)=0}^{\varphi(t)=0} {\cal D}\varphi(\tau) \, {\cal P}\bigl[\varphi\bigr]  
\prod_{a=1}^{N} 
\biggl[\int_{\phi_{a}(0)=0}^{\phi_{a}(t)=0} {\cal D}\phi_{a}(\tau) \biggr]
\exp\biggl\{
-\frac{1}{2} \beta \sum_{a=1}^{N} \int_{0}^{t} d\tau 
\Bigl[
\bigl(\partial_\tau \phi_{a}(\tau)\bigr)^2 + 
\lambda \bigl(\phi_{a}(\tau)-\varphi(\tau)\bigr)^{2}
\Bigr]
\biggr\}
\label{14}
\end{eqnarray}
is the replica partition function. Substituting here the probability distribution function
${\cal P}\bigl[\varphi\bigr]$, eq.(\ref{3}), we get
\begin{equation}
\label{15}
Z_{N} \; = \; \sqrt{\frac{2\pi t}{\gamma}} \,
\int_{\varphi(0)=0}^{\varphi(t)=0} {\cal D}\varphi(\tau)
\prod_{a=1}^{N} 
\biggl[\int_{\phi_{a}(0)=0}^{\phi_{a}(t)=0} {\cal D}\phi_{a}(\tau) \biggr]
\; 
\exp\Bigl\{ - {\cal H}_{N} [\phi_{a}; \varphi] \Bigr\}
\end{equation}
with the replica Hamiltonian
\begin{equation}
\label{16}
{\cal H}_{N} [\phi_{a}; \varphi] \, = \, 
\frac{1}{2} \int_{0}^{t} 
\Bigl[
\beta \sum_{a=1}^{N} \bigl(\partial_\tau \phi_{a}\bigr)^2 
+
\gamma \bigl(\partial_\tau \varphi\bigr)^2 
+
\beta \lambda \sum_{a=1}^{N} \bigl(\phi_{a}-\varphi\bigr)^{2}
\Bigr]
\end{equation}
To compute the replica partition function (\ref{15})
let us introduce more complicated function
\begin{equation}
\label{17}
\Psi[{\bf x}; \, y; \, t] = 
\int_{\varphi(0)=0}^{\varphi(t)=y} {\cal D}\varphi(\tau)
\prod_{a=1}^{N} 
\biggl[\int_{\phi_{a}(0)=0}^{\phi_{a}(t)=x_{a}} {\cal D}\phi_{a}(\tau) \biggr]
\; 
\exp\Bigl\{ - {\cal H}_{N} [\phi_{a}; \varphi] \Bigr\}
\end{equation}
(where ${\bf x} = \{x_{1}, x_{2}, ..., x_{N} \}$ ) such that
\begin{equation}
\label{18}
Z_{N} \; = \;  \sqrt{\frac{2\pi t}{\gamma}} \, \Psi[{\bf 0}; \, 0; \, t]
\end{equation}
One can easily show that the above function $\Psi[{\bf x}; \, y; \, t]$ satisfy the 
imaginary-time Schr\"odinger equation for (N+1) quantum bosons
\begin{equation}
\label{19}
\frac{\partial}{\partial t} \Psi \; = \; 
\frac{1}{2\beta} \sum_{a=1}^{N} \frac{\partial^{2}}{\partial x_{a}^{2}} \, \Psi 
\; + \; 
\frac{1}{2\gamma}  \frac{\partial^{2}}{\partial y^{2}} \, \Psi
\; - \; 
\frac{1}{2} \beta \lambda \, \sum_{a=1}^{N} (x_{a} - y)^{2} \, \Psi
\end{equation}
In the limit $t\to \infty$ the solution $\Psi[{\bf x}; \, y; \, t]$ of the above equation is dominated 
by the ground state wave function
\begin{equation}
\label{20}
\Psi[{\bf x}; \, y; \, t] \; \simeq \; \exp\bigl\{- E \, t\bigr\} \, \psi({\bf x}; \, y)
\end{equation}
As in this limit we will be interested only in the extensive part of the free energy,
in what follows all pre-exponetial factors (like $\sqrt{2\pi t/\gamma}$ in eq.(\ref{18})) will be omitted.
The ground state energy $E$ and the ground state eigenfunction 
$\psi({\bf x}; \, y)$ are defined by the equation
\begin{equation}
\label{21}
-E \, \psi \; = \; 
\frac{1}{2\beta} \sum_{a=1}^{N} \frac{\partial^{2}}{\partial x_{a}^{2}} \, \psi 
\; + \; 
\frac{1}{2\gamma}  \frac{\partial^{2}}{\partial y^{2}} \, \psi
\; - \; 
\frac{1}{2} \beta \lambda \, \sum_{a=1}^{N} (x_{a} - y)^{2} \, \psi
\end{equation}
It is evident that the solution of the above equation must have the Gaussian form
\begin{equation}
\label{22}
\psi({\bf x}; \, y) \; = \; 
\exp\biggl\{
-\frac{1}{2} A \sum_{a=1}^{N} x_{a}^{2} \; - \; \frac{1}{2} B \, y^{2} 
\; - \; C \Bigl(\sum_{a=1}^{N} x_{a}\Bigr) \, y \; - \; \frac{1}{2} D \Bigl(\sum_{a=1}^{N} x_{a}\Bigr)^{2}
\biggr\}
\end{equation}
where the parameters $A, \, B, \, C$ and $D$ are to be defined. Substituting eq.(\ref{22}) into
eq.(\ref{21}) we are getting the following set of equations
\begin{eqnarray}
\label{23}
&&
A^{2} \; = \; \beta^{2} \lambda
\\
\nonumber
\\
\label{24}
&&
\frac{1}{\gamma} B^{2} \, + \, \frac{N}{\beta} C^{2} \; = \; \beta \lambda \, N 
\\
\nonumber
\\
\label{25}
&&
\frac{N}{\beta} D^{2} \, + \, \frac{2}{\beta} A D \, + \, \frac{1}{\gamma} C^{2} \; = \; 0
\\
\nonumber
\\
\label{26}
&&
C \, \Bigl(\frac{1}{\beta} A \, + \, \frac{N}{\beta} D \, + \, \frac{1}{\gamma}B \Bigr) 
\; = \; - \beta \lambda 
\end{eqnarray}
and 
\begin{equation}
\label{27}
E \; = \; \frac{N}{2\beta} (A \, + \, D) \; + \; \frac{1}{2\gamma} B
\end{equation}
It can be easily shown that in the first order in $N \to 0$ 
the solution of these equations is
\begin{eqnarray}
\label{28}
A &=& \beta \sqrt{\lambda}
\\
\nonumber
\\
\label{29}
B &\simeq& \beta \sqrt{\lambda} \, N \; + \; o(N)
\\
\nonumber
\\
\label{30}
C &\simeq& -\beta \sqrt{\lambda}  \; + \; \frac{1}{2\gamma} \beta^{2} \sqrt{\lambda} \, N \; + \; o(N)
\\
\nonumber
\\
\label{31}
D &\simeq& - \frac{1}{2\gamma} \beta^{2} \sqrt{\lambda} 
\; + \; \frac{3}{8\gamma^{2}}\beta^{3}\sqrt{\lambda} \, N 
\; + \; o(N)
\end{eqnarray}
so that
\begin{equation}
\label{32}
E \; \simeq \;  
 N \, \Bigl( \frac{1}{2}\sqrt{\lambda} \; + \; \frac{\beta}{4\gamma}\sqrt{\lambda} \Bigr)
\end{equation}
Substituting the above results  into eqs.(\ref{20})  and  eq.(\ref{18}) 
(and neglecting pre-exponential factor) we get
\begin{equation}
\label{33}
Z_{N} \; \simeq \;  
\exp\biggl\{
- N \biggl( \frac{1}{2} \, \sqrt{\lambda} \; + \; \frac{\beta}{4\gamma}\sqrt{\lambda} \biggr) \, t 
\biggr\}
\end{equation}
Thus, according to eq.(\ref{13}), for the extensive in $t$ part of the free energy
one finally obtains
\begin{equation}
\label{34}
F(T, \lambda) \; = \; \biggl( \frac{1}{2} \, T \sqrt{\lambda} 
\; + \; \frac{1}{4\gamma}\sqrt{\lambda} \biggr) \, t 
\end{equation}
One can easily see that according to its definition, eq.(\ref{7}), the overlap parameter $q$ 
can be obtained as
\begin{equation}
\label{35}
q \; = \; \frac{2}{t} \, \frac{\partial}{\partial \lambda} \, F(T, \lambda)
\end{equation}
so that one immediately find the following result
\begin{equation}
\label{36}
q \; = \; \frac{1}{2\sqrt{\lambda}} 
\Bigl(
T \; + \; \frac{1}{2\gamma}
\Bigr)
\end{equation}
or, in original notations, eq.(\ref{12}),
\begin{equation}
\label{37}
q \; = \; \Bigl(\frac{\pi}{8}\Bigr)^{1/4} \,  
\Bigl(T + \frac{1}{2\gamma}\Bigr) \, \sqrt{\frac{\epsilon^{3}}{u} }
\end{equation}
Correspondingly, for the retrieval order parameter, eq.(\ref{8}), we get 
\begin{equation}
\label{38}
m(u, T) \; = \; \frac{\sqrt{u}}{\sqrt{u} \; + \; \bigl(\frac{\pi}{8}\bigr)^{1/4} \,  
	                                       \Bigl(T + \frac{1}{2\gamma}\Bigr) \epsilon^{3/2} }
\end{equation}

However, according to the discussion in the ending part of the Introduction,
the Gaussian approximation, eq.(\ref{9}), considered in this Section is valid 
only provided $q \ll \epsilon^{2}$. Using eq.(\ref{37}) we find that the above result 
for the retrieval parameter $m$ is valid only for 
\begin{equation}
\label{39}
\sqrt{u \, \epsilon} \; \gg \; \Bigl(T + \frac{1}{2\gamma}\Bigr) 
\end{equation}
We see that according to the result (\ref{38}),  
as the coupling parameter $u$ increases from it minimal allowed value 
$\sim \Bigl(T + \frac{1}{2\gamma}\Bigr)^{2}/\epsilon$ 
to infinity, the value of $m$ increases from a finite value to one 
(which corresponds to perfect retrieval).

\section{Short range interaction limit} 

In the short range interaction approximation, eq.(\ref{10}), 
the Hamiltonian of the considered system,
eq.(\ref{1}), becomes
\begin{equation}
\label{40}
H[\phi(\tau); \varphi(\tau)] = \int_{0}^{t} d\tau
\Bigl\{\frac{1}{2} \bigl(\partial_\tau \phi(\tau)\bigr)^2
\; - \; u \, \delta \bigl(\phi(\tau)-\varphi(\tau)\bigr)\Bigr\}
\end{equation}
Following the same rout of the replica formalism described in the previous Section,
instead of eq.(\ref{16}) for the replica Hamiltonian we get
\begin{equation}
\label{41}
{\cal H}_{N} [\phi_{a}; \varphi] \, = \, 
\frac{1}{2} \int_{0}^{t} 
\Bigl[
\beta \sum_{a=1}^{N} \bigl(\partial_\tau \phi_{a}\bigr)^2 
+
\gamma \bigl(\partial_\tau \varphi\bigr)^2 
+
2 \beta u  \sum_{a=1}^{N} \delta\bigl(\phi_{a}-\varphi\bigr)^{2}
\Bigr]
\end{equation}
Correspondingly, for the wave function (\ref{17}) (which defines the replica
partition function according to eq.(\ref{18})) we are getting the 
imaginary-time Schr\"odinger equation for (N+1) quantum bosons
\begin{equation}
\label{42}
\frac{\partial}{\partial t} \Psi({\bf x}, y, t) \; = \; 
\frac{1}{2\beta} \sum_{a=1}^{N} \frac{\partial^{2}}{\partial x_{a}^{2}} \, \Psi({\bf x}, y, t) 
\; + \; 
\frac{1}{2\gamma}  \frac{\partial^{2}}{\partial y^{2}} \, \Psi({\bf x}, y, t)
\; + \; 
 \beta u \, \sum_{a=1}^{N} \delta(x_{a} - y) \; \Psi({\bf x}, y, t)
\end{equation}
and in the representation (\ref{20}) we find that the ground state energy and the 
ground state wave function are defined by the equation
\begin{equation}
\label{43}
-E \, \psi({\bf x}, y) \; = \; 
\frac{1}{2\beta} \sum_{a=1}^{N} \frac{\partial^{2}}{\partial x_{a}^{2}} \, \psi ({\bf x}, y)
\; + \; 
\frac{1}{2\gamma}  \frac{\partial^{2}}{\partial y^{2}} \, \psi({\bf x}, y)
\; + \; 
 \beta u \, \sum_{a=1}^{N} \delta(x_{a} - y) \; \psi({\bf x}, y)
\end{equation}
The structure of the above equation implies that its solution 
is the function of the differences
$(x_{a} - y) = z_{a}$, namely, $\psi\bigl({\bf x},  y\bigr) \, = \, \psi\bigl({\bf z}\bigr)$.
In terms of these new variables instead of eq.(\ref{43}) we get
\begin{equation}
\label{44}
-E \, \psi({\bf z}) \; = \; 
\frac{1}{2\beta} \sum_{a=1}^{N} \frac{\partial^{2}}{\partial z_{a}^{2}} \, \psi ({\bf z})
\; + \; 
\frac{1}{2\gamma} \sum_{a,b=1}^{N} \frac{\partial^{2}}{\partial z_{a} \partial z_{b}} \, \psi({\bf z})
\; + \; 
\beta u \, \sum_{a=1}^{N} \delta(z_{a}) \; \psi({\bf z})
\end{equation}
Integrating this equation over a particular $z_{a}$ in a narrow interval around zero 
from $-\eta$ to $+\eta$ (where $\eta \ll 1$) and assuming that
$\psi({\bf z})$ is a continuous function, we find that the derivatives of this 
function has a singularity at $z_{a} = 0$:
\begin{equation}
\label{45}
\Bigl(\frac{1}{2\beta} \, + \, \frac{1}{2\gamma} \Bigr)
\biggl(
\frac{\partial}{\partial z_{a}} \psi({\bf z})\bigg|_{z_{a}=+\eta}  - \; 
\frac{\partial}{\partial z_{a}} \psi({\bf z})\bigg|_{z_{a}=-\eta}
\biggr)
\; = \; - \beta u \,\psi({\bf z})\Big|_{z_{a}=0}
\end{equation}
On the other hand, in the region of space where all $z_{a} \not= 0$ the function $\psi({\bf z})$
must satisfy the equation
\begin{equation}
\label{46} 
\frac{1}{2\beta} \sum_{a=1}^{N} \frac{\partial^{2}}{\partial z_{a}^{2}} \, \psi ({\bf z})
\; + \; 
\frac{1}{2\gamma} \sum_{a,b=1}^{N} \frac{\partial^{2}}{\partial z_{a} \partial z_{b}} \, \psi({\bf z})
\; = \; 
-E \, \psi({\bf z})
\end{equation}
One can easily show that for {\it negative} values of the energy $E$ 
the above equation has $2^{N}$ solutions labeled by $N$ Ising
parameters $\{\sigma_{a} = \pm 1 \} \; \; (a = 1, ..., N)$, such that a given set of $\sigma$'s
\begin{equation}
\label{47}
\psi({\bf z}) \; = \; 
\psi_{0} \, 
\exp
\biggl\{
-\omega \sum_{a=1}^{N} \sigma_{a} \, z_{a}
\biggr\}
\end{equation}
where $\psi_{0}$ is irrelevant normalization constant and
\begin{equation}
\label{48}
\omega \; = \; \frac{\sqrt{2|E|}}{\sqrt{\frac{1}{\beta} N \; + \; 
		                               \frac{1}{\gamma} \Bigl(\sum_{a=1}^{N} \sigma_{a}\Bigr)^{2} } }
\end{equation}
Let us consider a particular set $\{\sigma_{1}, \sigma_{2}, ..., \sigma_{N} \}$ 
in which $k$ elements are positive and $(N-k)$ elements are negative (where $k = 0, 1, ..., N$).
Reordering these  $\sigma$'s, this particular set can always be represented as
\begin{equation}
\label{49}
\sigma_{a} \; = \; \left\{
\begin{array}{ll}
+1, & \mbox{for $a = 1, 2, ..., k$} \\
\\
-1, & \mbox{for $a = k+1, ..., N$}
\end{array}
\right.
\end{equation}
In this case 
\begin{equation}
\label{50}
\sum_{a=1}^{N} \sigma_{a} = 2k - N
\end{equation}
Thus, for such particular choice  of $\sigma$'s, 
according to eq.(\ref{48}),
\begin{equation}
\label{51}
\omega \; = \; \frac{\sqrt{2|E|}}{\sqrt{\frac{1}{\beta} N \; + \; 
		\frac{1}{\gamma} (2k - N)^{2} } }
\end{equation}
As the solution of eq.(\ref{46}) must be symmetric with respect to permutations of $z_{a}$
it can be represented as follows:
\begin{equation}
\label{52}
\psi^{(+k)}({\bf z}) \; = \; 
\psi_{0} \, \sum_{{\cal P}^{k}_{N}} \exp
\biggl\{
-\omega \sum_{a=1}^{N} \sigma_{a} \, z_{{\cal P}^{k}_{N}(a)}
\biggr\} 
\; = \; 
\psi_{0} \, \sum_{{\cal P}^{k}_{N}} 
 \exp
\Bigl\{
-\omega \sum_{a=1}^{k} z_{{\cal P}^{k}_{N}(a)} + \omega \sum_{a=k+1}^{N} z_{{\cal P}^{k}_{N}(a)}
\Bigr\}
\end{equation}
where the summation goes over all permutations ${\cal P}^{k}_{N}$ of $N$ points $z_{a}$ over 
$(k \; + \; (N-k))$ elements $\sigma_{a}$ (which are defined in eq.(\ref{49})).
It is evident that for "antisymmetric" choice: $\sigma_{a} \to -\sigma_{a}$
(for which $\sum_{a=1}^{N} \sigma_{a} = -(2k - N)$) the function
\begin{equation}
\label{53}
\psi^{(-k)}({\bf z}) \; = \; 
\psi_{0} \, \sum_{{\cal P}^{k}_{N}} 
\exp
\Bigl\{
\omega \sum_{a=1}^{k} z_{{\cal P}^{k}_{N}(a)} - \omega \sum_{a=k+1}^{N} z_{{\cal P}^{k}_{N}(a)}
\Bigr\}
\end{equation}
is also the solution of eq.(\ref{46}). Let us remind now that
the solution of the original equation (\ref{44}) must satisfy the conditions (\ref{45}).
Taking into account these conditions, 
one can easily see that for a given permutation ${\cal P}^{k}_{N}$ the function which satisfy 
both eqs.(\ref{45}) and eq.(\ref{46}), and decaying zero for all 
$z_{a} \to \pm\infty$ can be defined as follows
\begin{equation}
\label{54}
\psi^{(k)}_{{\cal P}^{k}_{N}}({\bf z})  =  \left\{
\begin{array}{ll}
 \exp
\Bigl\{
-\omega \sum_{a=1}^{k} z_{{\cal P}^{k}_{N}(a)} + \omega \sum_{a=k+1}^{N} z_{{\cal P}^{k}_{N}(a)}
\Bigr\}, & \mbox{for $\{z_{{\cal P}^{k}_{N}(1)}, ..., z_{{\cal P}^{k}_{N}(k)}\} > 0$ 
	         and  $\{z_{{\cal P}^{k}_{N}(k+1)}, ..., z_{{\cal P}^{k}_{N}(N)}\} < 0$} \\
\\
\\
 \exp
\Bigl\{
+\omega \sum_{a=1}^{k} z_{{\cal P}^{k}_{N}(a)} - \omega \sum_{a=k+1}^{N} z_{{\cal P}^{k}_{N}(a)}
\Bigr\}, & \mbox{for $\{z_{{\cal P}^{k}_{N}(1)}, ..., z_{{\cal P}^{k}_{N}(k)}\} < 0$ 
	and  $\{z_{{\cal P}^{k}_{N}(k+1)}, ..., z_{{\cal P}^{k}_{N}(N)}\} > 0$}
\end{array}
\right.
\end{equation}
where, according to eq.(\ref{45}), the parameter $\omega$ must satisfy
the condition
\begin{equation}
\label{55}
\Bigl(\frac{1}{2\beta} \, + \, \frac{1}{2\gamma} \Bigr)
\bigl(-2\omega \bigr)
\; = \; - \beta u \,\psi({\bf z})\Big|_{z_{a}=0}
\end{equation}
or
\begin{equation}
\label{56}
\omega
\; = \; 
\frac{\beta^{2}  u \gamma}{(\beta + \gamma)}
\end{equation}
Comparing this value of $\omega$ with the one in eq.(\ref{51}) we find the corresponding eigenenergy:
\begin{equation}
\label{57}
E_{k} \; = \; 
- \frac{\beta^{4} u^{2} \gamma^{2}}{2 (\beta + \gamma)^{2}} \, 
\Bigl(\frac{1}{\beta} N \; + \; \frac{1}{\gamma} (2k - N)^{2}\Bigr)
\end{equation}
Thus we conclude that eq.(\ref{44}) exhibits the {\it spectrum} of solutions
labeled by the integer parameter $k \; = \; 0, 1, 2, ..., N$:
\begin{equation}
\label{58}
\psi^{(k)}({\bf z}) \; = \; 
\psi_{0} \, \sum_{{\cal P}^{k}_{N}} \, \psi^{(k)}_{{\cal P}^{k}_{N}}({\bf z})
\end{equation}
where the function $\psi^{(k)}_{{\cal P}^{k}_{N}}({\bf z})$ is given in eq.(\ref{54}), the parameter $\omega$
is given in eq.(\ref{56}) and the eigenenergy $E_{k}$ is given in eq.(\ref{57}).
Correspondingly, according to eqs.(\ref{18}) and (\ref{20}), for the replica partition function 
(neglecting pre-exponential factor) we get
\begin{equation}
\label{59}
Z_{N} \; \sim \; \sum_{k=0}^{N} \; \exp\bigl\{-E_{k} t \bigr\}
 \sum_{{\cal P}^{k}_{N}} \, \psi^{(k)}_{{\cal P}^{k}_{N}}({\bf 0}) 
 \; = \; 
\sum_{k=0}^{N} \frac{N!}{k! \, (N-k)!} 
\exp
\biggl\{
\frac{\beta^{4} u^{2} \gamma^{2}}{2 (\beta + \gamma)^{2}} \, 
\Bigl(\frac{1}{\beta} N \; + \; \frac{1}{\gamma} (2k - N)^{2}\Bigr) \; t
\biggr\}
\end{equation}
Simple calculations yield:
\begin{equation}
\label{60}
Z_{N} \; \sim \;
\exp
\biggl\{
\frac{\beta^{2} u^{2} \gamma^{2}}{2 (\beta + \gamma)^{2}} \,  (\beta N) \, t
\biggr\}
\times S
\end{equation}
where
\begin{eqnarray}
\nonumber
S  &=& \sum_{k=0}^{N} \frac{N!}{k! \, (N-k)!} 
\exp
\biggl\{
\frac{\beta^{4} u^{2} \gamma}{2 (\beta + \gamma)^{2}} \,  (2k-N)^{2} \, t
\biggr\}
\\
\nonumber
\\
\nonumber
&=&
\int_{-\infty}^{+\infty} \frac{d\xi}{\sqrt{2\pi}} \exp\Bigl\{-\frac{1}{2} \xi^{2} \Bigr\} \;
\sum_{k=0}^{N} \frac{N!}{k! \, (N-k)!}
\exp
\biggl\{
\frac{\beta^{2} u \sqrt{\gamma}}{(\beta + \gamma)} \,  (2k-N)\sqrt{t} \; \xi
\biggr\}
\\
\nonumber
\\
&=&
\int_{-\infty}^{+\infty} \frac{d\xi}{\sqrt{2\pi}} \exp\Bigl\{-\frac{1}{2} \xi^{2} \Bigr\} \;
\biggl[
2 \, \cosh\Bigl(
\frac{\beta^{2} u \sqrt{\gamma}}{(\beta + \gamma)}\sqrt{t} \; \xi
\Bigr)
\biggr]^{N}
\label{61}
\end{eqnarray}
In the limit $t\to\infty$,
\begin{equation}
\label{62}
S \; \simeq \; 
\int_{-\infty}^{+\infty} \frac{d\xi}{\sqrt{2\pi}} 
\exp
\Bigl\{-\frac{1}{2} \xi^{2} 
\; + \; 
\frac{\beta^{2} u \sqrt{\gamma}}{(\beta + \gamma)}\sqrt{t} \; N \; |\xi|
\Bigr\} 
\; \sim \; 
\exp
\biggl\{
\frac{\beta^{2} u^{2} \gamma}{2 (\beta + \gamma)^{2}} \, (\beta N)^{2} \, t
\biggr\}
\end{equation}
Thus, for the replica partition function (\ref{60}) (in the limit $t\to\infty$) we get the following result
\begin{equation}
\label{63}
Z_{N} \; \sim \;
\exp
\biggl\{
\frac{\beta^{2} u^{2} \gamma^{2}}{2 (\beta + \gamma)^{2}} \,  (\beta N) \, t
\; + \; 
\frac{\beta^{2} u^{2} \gamma}{2 (\beta + \gamma)^{2}} \, (\beta N)^{2} \, t
\biggr\}
\end{equation}

As the partition function of a system and its and free energy are related as 
$Z = \exp\{-\beta F\}$, according to its definition, the replica partition function 
$Z_{N} \; = \; \overline{Z^{N}}$
is related with the free energy distribution function $P(F)$ as follows
\begin{equation}
\label{64}
Z_{N}\; =\;
\int_{-\infty}^{+\infty} dF \, P(F) \;
\exp\bigl\{ -\beta N F\bigr\}
\end{equation}
According to eq.(\ref{63}), we have
\begin{equation}
\label{65}
\int_{-\infty}^{+\infty} dF \, P(F) \;
\exp\bigl\{ -(\beta N) F\bigr\}
\; = \; 
\exp
\biggl\{
\frac{\beta^{2} u^{2} \gamma^{2}}{2 (\beta + \gamma)^{2}} \,  (\beta N) \, t
\; + \; 
\frac{\beta^{2} u^{2} \gamma}{2 (\beta + \gamma)^{2}} \, (\beta N)^{2} \, t
\biggr\}
\end{equation}
Using the above relation one easily finds that 
\begin{equation}
\label{66}
P(F) \; = \;\frac{1}{\sqrt{2\pi} (\delta F)} \,
\exp\Biggl\{-\frac{\bigl(F - \overline{F}\bigr)^{2}}{2 (\delta F)^{2}}\Biggr\}.
\end{equation}
where 
\begin{equation}
\label{67}
\overline{F} \; = \; 
\frac{\beta^{2} u^{2} \gamma^{2}}{2 (\beta + \gamma)^{2}} \,  t
\end{equation}
is the average free energy and
\begin{equation}
\label{68}
\delta F \; \equiv \; \sqrt{\overline{\bigl(F - \overline{F}\bigr)^{2}} } 
\; = \;  
\frac{\beta u \sqrt{\gamma}}{(\beta + \gamma)} \;  t^{1/2}
\end{equation}
is the mean square fluctuation of the free energy.

According to the definition, eq.(\ref{7}), the overlap parameter $q$ can be estimated
as $q \, \sim \, \langle z^{2}\rangle$, where, according to eqs.(\ref{54}) and (\ref{56}),
$\langle z^{2}\rangle \, \sim \, 1/\omega^{2}$, or
\begin{equation}
\label{69}
q \; \sim \; \frac{(\beta + \gamma)^{2}}{\beta^{4} u^{2} \gamma^{2}} \; = \; 
\frac{1}{u^{2}} \, T^{2} \Bigl(T \, + \, \frac{1}{\gamma}\Bigr)^{2}
\end{equation}
Correspondingly, for the retrieval order parameter, eq.(\ref{8}), we get 
\begin{equation}
\label{70}
m(u, T) \; \sim \; \frac{u^{2}}{u^{2} \; + \;  T^{2} \Bigl(T \, + \, \frac{1}{\gamma}\Bigr)^{2}}
\end{equation}
Note that the short range approximation, eq.(\ref{10}), considered in this Section is valid only
for $q \gg \epsilon^{2}$,  or for  
\begin{equation}
\label{71}
\epsilon u \; \ll \;  T \Bigl(T \, + \, \frac{1}{\gamma}\Bigr)
\end{equation}

\section{Discussion}

In conclusion, in this paper we have proposed very simple statistical "memory model" 
of one-dimensional directed polymers which is capable to store and retrieve 
a given random quenched trajectory. The model is defined in terms of the 
elastic string Hamiltonian (\ref{1}) with the local attractive potential between
the dynamic and the quenched random strings. We have calculated the average 
overlap between the dynamic and the quenched string $q(T, u)$, eq.(\ref{7}), 
as well as the effective "retrieval parameter" $m(T, u)$, eq.(\ref{8}). 
The explicit expressions for these parameters as the functions of the 
temperature $T$ and the coupling constant $u$ of the attractive potential $U(\phi)$
are calculated in the two limit cases: first, in the harmonic approximation of the 
potential $U(\phi)$, eq.(\ref{9}), and second, in the $\delta$-function approximation
of $U(\phi)$, eq.(\ref{10}). It is shown that $m(T, u)$ is the smooth function of 
$T$ and $u$, ranging from one (perfect retrieval) to zero (no retrieval).
Summarising the results derived in Section II and III, eqs.(\ref{39})-(\ref{38}), 
(\ref{70})-(\ref{71}), for the "retrieval parameter" $m(u,T)$
we get
\begin{equation}
\label{72}
m(u, T)  \; \sim \;   \left\{
\begin{array}{ll}
\frac{(\epsilon u)^{2}}{(\epsilon u)^{2} \; + \;  
	T^{2} \Bigl(T \, + \, \frac{1}{\gamma}\Bigr)^{2} \epsilon^{2}} = m_{\delta}(u, T), & 
\mbox{for $u \; \ll \; \frac{1}{\epsilon} \, T \Bigl(T \, + \, \frac{1}{\gamma}\Bigr) 
	\, = \, u_{\delta}(T) $} 
\\
\\
\frac{\sqrt{\epsilon u}}{\sqrt{\epsilon u} 
	\; + \; \bigl(\frac{\pi}{8}\bigr)^{1/4} \, \Bigl(T + \frac{1}{2\gamma}\Bigr) \epsilon^{2} }
 = m_{g}(u, T), & 
\mbox{for $ u \; \gg \;  \frac{1}{\epsilon} \, \Bigl(T \, + \, \frac{1}{2\gamma}\Bigr)^{2}
	\, = \, u_{g}(T)	$}
\end{array}
\right.
\end{equation}
where $\epsilon$ is the spatial size of the potential $U(\phi)$, eq.(\ref{2}), 
and $\gamma$ is the elastic parameter of quenched random string, eq.(\ref{3}). 
Note that at the crossover between the  two regimes, the values of $m(u,T)$ 
nicely fits with each other being independent of $T$ and $u$:
\begin{equation}
\label{73}
m_{g}(u, T)\Big|_{ u = u_{g}(T)} \, = \, 
\frac{1}{1 \; + \; \bigl(\frac{\pi}{8}\bigr)^{1/4} \,\epsilon^{2}}
\; \sim \; 
\frac{1}{1 \; + \; \epsilon^{2}}
\; \sim \; 
m_{\delta}(u, T)\Big|_{u = u_{\delta}(T)}
\end{equation}
The examples of the curves illustrating the dependence of the retrieval parameter $m(u, T)$ 
on the coupling constant $u$ are shown in Figure 1.
\begin{figure}[h]
	\begin{center}
		\includegraphics[width=12.0cm]{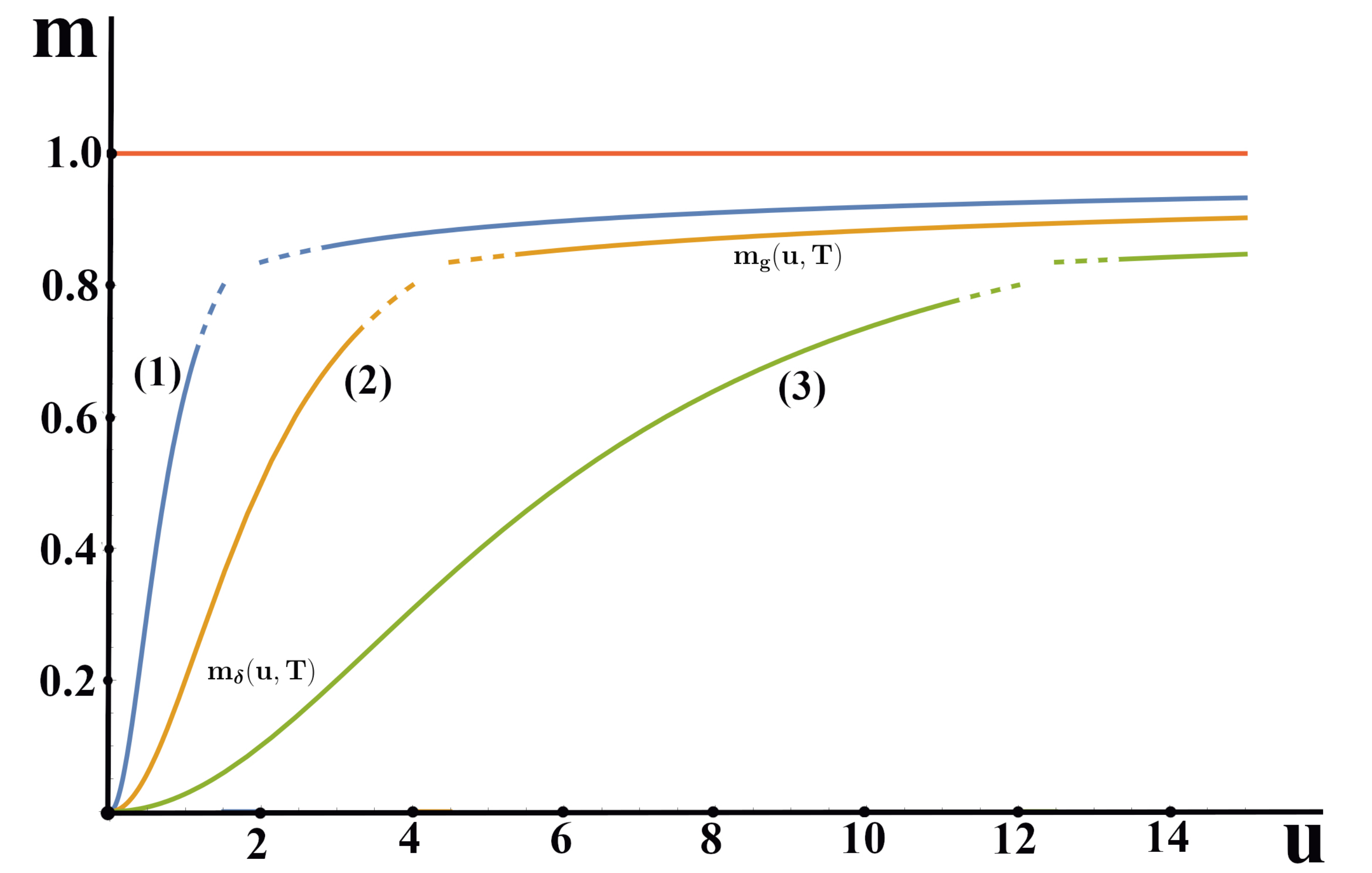}
		\caption[]{The retrieval parameter $m(u,T)$, eq.(\ref{72}), as a function of the 
		strength $u$ of the attractive potential $U(\phi)$, eq.(\ref{2}),  for three
	different temperatures: (1) $T=0.5$, (2) $T=1$ and (3) $T=2$, with $\epsilon = 0.5$ (size of the potential $U(\phi)$) and $\gamma =1$ (elastic parameter of a quenched random string) }
	\end{center}
	\label{figure1}
\end{figure}

Unfortunately, the generalization of the considered model for storing and retrieving of several 
quenched trajectories  (similar to storing of several patterns in neural networks) 
looks rather problematic. The problem is that the trajectories with {\it independent} 
distributions will be inevitably intersecting, which makes the identification of a given trajectory
ambiguous. On the other hand, one can of course forbid the intersection of the quenched trajectories
(e.g. by introduction $\delta$-like repulsion in their {\it joint} distribution function),
but as a consequence one will face much more sophisticated calculations (compared to the ones of the 
present paper) which still remains to be done.

%%%%%%%%%%%%%%%%%%%%%%%%%%%%%%%%%%%%%%%%%


\begin{thebibliography}{99}
	
%%%%%%%%%%%%% Introduction

\bibitem{KPZ} M.Kardar, G.Parisi,Y-C.Zhang,
     Phys.\ Rev.\ Lett.\ {\bf 56}, 889 (1986)

\bibitem{hh_zhang_95} T.\ Halpin-Healy and Y-C.\ Zhang,
    Phys.\ Rep.\ {\bf 254}, 215 (1995)

\bibitem{Corwin} I.\ Corwin,
    Random Matrices: Theory Appl. {\bf 1}, 1130001 (2012)

\bibitem{Borodin} A.\ Borodin, I.\ Corwin and P.\ Ferrari,
    Comm. Pure Appl. Math. {\bf 67}, 1129–1214 (2014)

\bibitem{Rev} V.Dotsenko,
    {\it Universal Randomness}, Physics-Uspekhi, {\bf 54}(3), 259 (2011);\\
    {\it Statistical properties of one-dimensional directed polymers in a random potential},
    in {\it Order, Disorder and Criticality}, vol 5 (World Scientific, 2018); arXiv:1703.04305

%%%%%%%%%%%%%%

\bibitem{Amit} D. Amit, H. Sompolinsky and H. Gutfreund, 
   Ann.Phys. {\bf 173} 30 (1987)

\bibitem{book1} V.S.Dotsenko, 
{\it Introduction to the Theory of Spin Glasses and Neural Networks} (World Scientific, 1994)


\end{thebibliography}
\end{document}